\documentclass[12pt]{article}
\usepackage{epsfig,amsmath,amsfonts,amssymb,amstext,afterpage,psfrag,
}
\allowdisplaybreaks
\begin{document}

\begin{titlepage}

\begin{flushright}
{\small
}
\end{flushright}

\vspace{0.5cm}
\begin{center}
{\Large\bf \boldmath                                               
Closing in on the radiative weak chiral couplings 
\vspace*{0.3cm}                                                  
\unboldmath}
\end{center}

\vspace{0.5cm}
\begin{center}
{\sc Luigi Cappiello$^{1,2}$, Oscar Cat\`a$^{3}$ and Giancarlo D'Ambrosio$^{2}$} 
\end{center}

\vspace*{0.4cm}

\begin{center}
$^1$ {\small{Dipartimento di Scienze Fisiche, Universit\'a di Napoli 'Federico II',}}\\[-0.05cm]
{\small{Via Cintia, 80126 Napoli, Italy}}\\[1.2mm]
$^2$ {\small{INFN-Sezione di Napoli, Complesso Universitario di Monte S. Angelo,}}\\[-0.05cm]
{\small{Via Cintia Edificio 6, 80126 Napoli, Italy}}\\[1.2mm]
$^3$ {\small{Theoretische Physik 1, Universit\"at Siegen, Walter-Flex-Stra\ss e 3, D-57068 Siegen, Germany}}\\[1.2mm]
\end{center}

\vspace{1cm}
\begin{abstract}
\vspace{0.2cm}
\noindent
We point out that, given the current experimental status of radiative kaon decays, a subclass of the ${\cal O} (p^4)$ counterterms of the weak chiral lagrangian can be determined in closed form. This involves in a decisive way the decay $K^\pm \to \pi ^\pm  \pi ^0 l^+  l^-$, currently being measured at CERN by the NA48/2 and NA62 collaborations. We show that consistency with other radiative kaon decay measurements leads to a rather clean prediction for the ${\cal{O}}(p^4)$ weak couplings entering this decay mode. This results in a characteristic pattern for the interference Dalitz plot, susceptible to be tested already with the limited statistics available at NA48/2. We also provide the first analysis of $K_S\to \pi^+\pi^-\gamma^*$, which will be measured by LHCb and will help reduce (together with the related $K_L$ decay) the experimental uncertainty on the radiative weak chiral couplings. A precise experimental determination of the ${\cal{O}}(p^4)$ weak couplings is important in order to assess the validity of the existing theoretical models in a conclusive way. We briefly comment on the current theoretical situation and discuss the merits of the different theoretical approaches. 
\end{abstract}

\vfill

\end{titlepage}

\section{Introduction}
\label{sec:intro}

Radiative kaon decays have been, and still are, one of the most powerful tools we have to explore the physics behind $\Delta S=1$ processes, most prominently by performing chiral tests and studies on CP violation. Chiral tests are a direct probe of the Standard Model at low energies. In this regime, physics is described by chiral perturbation theory (ChPT), where nonperturbative effects are parametrized in terms of a set of low-energy couplings. Keeping only the octet contributions, the $\Delta S=1$ Lagrangian up to NLO order is given by 
\begin{align}
{\cal{L}}_{\Delta S=1}=G_8 f_{\pi}^4~{\mathrm{tr}}\big[\Delta u_{\mu}u^{\mu}\big]+G_8f_{\pi}^2\sum_jN_j W_j+{\cal{O}}(p^6)~,
\end{align}
where $u^2=U$ is the chiral matrix containing pions and kaons, $u_{\mu}=iu^{\dagger}D_{\mu}U u^{\dagger}$ with $D_{\mu}U=\partial_{\mu}U+ieA_{\mu}[Q,U]$, $Q=\tfrac{1}{3}{\mathrm{diag}}[2,-1,-1]$ is the electric charge matrix and $\Delta=u\lambda_6u^{\dagger}$ is the chiral spurion for the $\Delta S=1$ transitions. The second term collects the NLO operators $W_j$. Out of the 37 operators present, only combinations of $W_{14}-W_{18}$ and $W_{28}-W_{31}$ are relevant for radiative kaon decays.  

As opposed to their analogs in the bottom and charm sector, radiative kaon decays are generically dominated by Bremsstrahlung, while resonance effects are suppressed. This makes the extraction of the weak counterterms quite challenging. The best strategy to determine the chiral couplings is to study the interference term between the Bremsstrahlung and resonance contributions, which is substantially bigger than the resonance piece. Since the interference term is linear in the counterterms, this has the additional advantage that one is sensitive to both their magnitude and sign. 

This strategy has been applied, e.g., to the decay modes $K^{\pm}\to \pi^{\pm}\gamma^*$, $K_S\to \pi^0\gamma^*$, $K^{\pm}\to \pi^{\pm}\pi^0\gamma$ or $K^{\pm}\to \pi^{\pm}\gamma\gamma$. This way three independent combinations of counterterms are known. In order to fully determine the set of weak counterterms $N_{14}-N_{18}$ an extra decay mode has to be measured. From this perspective, $K^+\to \pi^+\pi^0\gamma^*$ stands out as the most promising candidate: besides counterterm combinations already present in $K^{\pm}\to \pi^{\pm}\gamma^*$ and $K^{\pm}\to \pi^{\pm}\pi^0\gamma$, it contains an additional combination. Furthermore, it has already been measured by NA48/2, which collected of the order of $5000$ events thereof, statistics that will be improved by the new data currently being collected at NA62. 

The main observation of this note is that the undetermined counterterm in $K^+\to \pi^+\pi^0\gamma^*$ turns out to be sizeable. Our observation is not based on model estimates, but arises from combining the different experimental results on radiative kaon decays. If this expectation on the size of the counterterm is combined with the strategy laid out in~\cite{Pichl:2000ab,Cappiello:2011qc}, which shows how cuts in the dilepton invariant mass can diminish the dominance of Bremsstrahlung effects, a measurement could be performed even with the NA48/2 data. In particular, we provide a detailed analysis of the interference term to aid the counterterm determination. 

In the last part of the paper we comment on the decay mode $K_S\to \pi^+\pi^- e^+e^-$. In spite of the fact that its measurement is rather challenging, there are prospects to measure it at LHCb in the near future. For this decay mode the relevant counterterm combinations turn out to be entirely predicted from the measurements on other radiative kaon decay modes, so that one can have a rather precise estimate of how the interference term should look. LHCb could therefore provide a very relevant consistency check of the counterterm structure, and potentially improve the precision of the weak chiral couplings.

We conclude this note with some remarks on the present theoretical understanding of the radiative weak chiral couplings.

\section{Experimental status of the radiative chiral weak counterterms}

The radiative kaon decay modes $K^{\pm}\to \pi^{\pm}\gamma^*$, $K_S\to \pi^0\gamma^*$, $K^{\pm}\to \pi^{\pm}\pi^0\gamma$ and $K^+\to \pi^+\gamma\gamma$ have been observed and measured by the NA48 collaboration over the last 15 years. Besides the determination of the branching ratios, an extraction of the relevant combinations of chiral weak counterterms for those decays has also been possible. The chiral counterterms are related to the slope of the differential decay rate, which can be most easily accessed through the interference term between Bremsstrahlung and electric emission. Using this technique, the experimental values for the slopes of the different modes were found to be
\begin{align}
&K^{\pm}\to \pi^{\pm}\gamma^*:& &a_+=-0.578\pm 0.016~\cite{Batley:2009aa,Batley:2011zz}\\
&K_S\to \pi^0\gamma^*:& &a_S=(1.06^{+0.26}_{-0.21}\pm 0.07)~\cite{Batley:2003mu,Batley:2004wg}\\
&K^{\pm}\to \pi^{\pm}\pi^0\gamma:& &X_E=(-24\pm 4\pm 4)~{\mathrm{GeV}}^{-4}~\cite{Batley:2010aa}\\
&K^+\to \pi^+\gamma\gamma:& &{\hat{c}}=1.56\pm 0.23\pm 0.11~\cite{Ceccucci:2014oza}~.
\end{align}
The results are linked to the different weak counterterm combinations through (see~\cite{Ecker:1987qi,DAmbrosio:1998gur,Cappiello:2007rs,DAmbrosio:1996cak})
\begin{align}
{\cal{N}}_E^{(1)}&\equiv N_{14}^r-N_{15}^r=\frac{3}{64\pi^2}\left(\frac{1}{3}-\frac{G_F}{G_8}a_+-\frac{1}{3}\log\frac{\mu^2}{m_K m_{\pi}}\right)-3L_9^r~;\\
{\cal{N}}_S&\equiv 2N_{14}^r+N_{15}^r=\frac{3}{32\pi^2}\left(\frac{1}{3}+\frac{G_F}{G_8}a_S-\frac{1}{3}\log\frac{\mu^2}{m_K^2}\right)~;\\
{\cal{N}}_E^{(0)}&\equiv N_{14}^r-N_{15}^r-N_{16}^r-N_{17}=-\frac{|{\cal{M}}_K| f_{\pi}}{2G_8}X_E~;\\
{\cal{N}}_0&\equiv N_{14}^r-N_{15}^r-2N_{18}^r=\frac{3}{128\pi^2}{\hat{c}}-3(L_9^r+L_{10}^r)~,
\end{align} 
where $|{\cal{M}}_K|=1.81\times 10^{-8}$ GeV is the amplitude for $K^+\to \pi^+\pi^0$, $f_{\pi}=93$ MeV is the pion decay constant and $G_8=9.1\times 10^{-6}$ GeV$^{-2}$. The results for the counterterms are summarized in Table~\ref{tab:1}, where we have used the values
\begin{align}
L_9^r(m_{\rho})=(5.9\pm 0.4)\cdot 10^{-3};\qquad L_{10}^r(m_{\rho})=(-4.1\pm 0.4)\cdot 10^{-3}~,
\end{align}
taken from~\cite{Bijnens:2002hp} and \cite{Rodriguez-Sanchez:2016jvw}, respectively. We remark that while ${\cal{N}}_E^{(0)}$ and ${\cal{N}}_0$ are scale-independent, ${\cal{N}}_E^{(1)}$ and ${\cal{N}}_S$ depend on scale. All quantities have been evaluated at the kaon mass scale.

\begin{table}[t]
\begin{center}
 \begin{tabular}{c|c|c } 
\mbox{Decay mode} & \mbox{counterterm combination}&\mbox{expt. value}\\ \hline
$K^\pm\to \pi ^{\pm} \gamma^*$&$N_{14}-N_{15}$ &$-0.0167(13)$\\ \hline
$K_{S}\to \pi ^{0} \gamma^*$&$2 N_{14}+N_{15} $&$+0.016(4)$\\ \hline
$K^{\pm }\to \pi ^{\pm }\pi ^{0}\gamma$&$N_{14}-N_{15}-N_{16}-N_{17}$&$+0.0022(7)$\\ \hline
$K^{\pm }\to \pi ^{\pm}\gamma\gamma  $ &$N_{14}-N_{15}-2N_{18}$&$-0.0017(32)$\\
\end{tabular} 
\end{center}
\caption{{\footnotesize{\it{Values of the counterterm combinations together with the decay mode from which they can most precisely be extracted.}}}}\label{tab:1}
\end{table}
\section{Prospects for $K^+\to \pi^+\pi^0e^+e^-$} 

NA48/2 has collected roughly $5\times 10^3$ $K^+\to \pi^+\pi^0e^+e^-$ events~\cite{Bloch-Devaux:2017goa}. With this rather limited statistics one is experimentally sensitive to the Bremsstrahlung and magnetic terms, which are known from $K^+\to \pi^+\pi^0$ (through Low's theorem) and from $K^+\to \pi^+\pi^0\gamma$, respectively. Whether one can also extract information on the (much suppressed) electric counterterms depends on the size of the interference term. Besides ${\cal{N}}_E^{(0)}$ and ${\cal{N}}_E^{(1)}$, defined in the previous section, $K^+\to \pi^+\pi^0e^+e^-$ also contains the counterterm combination
\begin{align}\label{weakN}
{\cal{N}}_E^{(2)}&=N_{14}^r+2N_{15}^r-3(N_{16}^r-N_{17}),
\end{align}
which is so far undetermined. The interplay of the three counterterms will determine whether NA48/2 is sensitive to the chiral counterterms. In this section we will show that ${\cal{N}}_E^{(2)}$ is expected to be positive and sizeable. This expectation is mostly determined by consistency with the experimental values of the previous section and is rather robust. We will then show how ${\cal{N}}_E^{(0)}$, ${\cal{N}}_E^{(1)}$ and the estimated ${\cal{N}}_E^{(2)}$ define a rather characteristic pattern for the interference Dalitz plot, which can be tested by using cuts on the invariant dilepton mass.
 
The decay $K^+\to \pi^+\pi^0e^+e^-$ was first studied in~\cite{Pichl:2000ab,Cappiello:2011qc}. The amplitude for this process can be factorized into leptonic and hadronic pieces as 
\begin{align}
{\cal{M}}=\frac{e}{q^2}[{\bar{u}}\gamma^{\mu}v]H_{\mu}(p_+,p_0,q)~,
\end{align}
where 
\begin{align}
H^{\mu}(p_+,p_0,q)=F_1p_+^{\mu}+F_2 p_0^{\mu}+F_3\epsilon^{\mu\nu\lambda\rho}p_{+\nu}p_{0\lambda}q_{\rho}~.
\end{align}
The last form factor contains the magnetic emission, already measured in $K^+\to \pi^+\pi^0\gamma$~\cite{Batley:2010aa}. An independent measurement through $K^+\to \pi^+\pi^0e^+e^-$ would be interesting, and a detailed analysis on how to do that was already presented in~\cite{Cappiello:2011qc}. Here we are mostly concerned with the weak chiral couplings associated with the electric emission. They enter the form factors as
\begin{align}\label{resKplus}
F_1&=\frac{2ie}{2q\cdot p_K-q^2}\frac{2q\cdot p_0}{2q\cdot p_++q^2}{\cal{M}}_Ke^{i\delta_0^2}-\frac{2ie G_8e^{i\delta_1^1}}{f_{\pi}}\left\{ q\cdot p_0{\cal{N}}_E^{(0)}+\frac{2}{3} q^2\bigg({\cal{N}}_E^{(1)}+3L_9\bigg)\right\},\nonumber\\
F_2&=-\frac{2ie}{2q\cdot p_K-q^2}{\cal{M}}_Ke^{i\delta_0^2}+\frac{2ie G_8e^{i\delta_1^1}}{f_{\pi}}\left\{q\cdot p_+{\cal{N}}_E^{(0)}-\frac{1}{3} q^2{\cal{N}}_E^{(2)}\right\},
\end{align}
where $\delta_0^2$ and $\delta_1^1$ are strong phases associated with final state interactions and ${\cal{M}}_K$ is the amplitude for the $K^+\to \pi^+\pi^0$ decay. The counterterm combinations ${\cal{N}}_E^{(0)}$ and ${\cal{N}}_E^{(1)}$ have already been determined in the previous section, while ${\cal{N}}_E^{(2)}$ is defined in eq.~(\ref{weakN}) above. Both ${\cal{N}}_E^{(1)}+3L_9$ and ${\cal{N}}_E^{(2)}$ depend on scale. However, in both cases this dependence turns out to be very mild and well within the experimental errors. We will accordingly omit it, as was done in~\cite{Cappiello:2011qc}. 

It should be noted that eq.~(\ref{resKplus}) differs from a NLO computation using chiral perturbation theory (see e.g.~\cite{Pichl:2000ab}): the Bremsstrahlung contribution is here evaluated using Low's theorem, which is an exact current algebra result, whereas the weak and strong countertems are evaluated at the classical level, neglecting loop corrections. Since ${\cal{M}}_K$ is taken from experiment, we are including the effects of final state interactions (pion rescattering), which are known to be important.

The decay rate is overwhelmingly dominated by the Bremsstrahlung component. However, as initially pointed out in~\cite{Pichl:2000ab}, this dominance decreases as the dilepton invariant energy $q^2$ is increased. Judicious cuts in $q^2$ can therefore allow experimental analyses to reach information on the weak counterterms. This strategy was developed in detail in~\cite{Cappiello:2011qc}, suggesting an optimal cut at $q_c\sim 50$ MeV.\footnote{$q_c$ is defined as the lower integration limit of the integral in $q$. Accordingly, $q_c=\sqrt{2}m_e$ corresponds to performing the whole integral in $q$, while increasing $q_c$ progressively eliminates the low-$q$ region, where the Bremsstrahlung dominates.}

The analysis done in~\cite{Cappiello:2011qc} was mostly meant for a first experimental analysis of {\mbox{$K^+\to \pi^+\pi^0e^+e^-$}}, with the determination of the Bremsstrahlung contribution and the magnetic piece as main goals (see~\cite{Bloch-Devaux:2017goa}). At the time, the electric contribution seemed inaccessible without NA62 data. Accordingly, sample values were chosen for the counterterms, mostly to show the shape of the Dalitz plot and get an order of magnitude estimate for the electric contributions. Presently there is an active effort to study the electric counterterms with NA48/2 data. This motivates to refine the analysis presented in~\cite{Cappiello:2011qc}.  

The relevant piece for the extraction of the weak chiral couplings is the interference term. Integrating over the angular variables, the differential decay rate can be parametrized as  
\begin{align}
\frac{d^3\Gamma_{\textsc{Int}}}{dE_{\gamma} d{\hat{T}}_c dq^2}&=\frac{\alpha^2 G_8\cos\delta}{72\pi^3f_{\pi}m_K^2 q^4}\frac{2m_l^2+q^2}{(2E_{\gamma}m_K-q^2)(2E_{\gamma}-m_K+2{\hat{T}}_c)}|{\cal{M}}_{K}|\sqrt{1-\frac{4m_l^2}{q^2}}\nonumber\\
&\!\!\!\!\!\times\Big[{\cal{N}}_E^{(0)}f_0(E_{\gamma},{\hat{T}}_c,q^2)+2q^2({\cal{N}}_E^{(1)}+3L_9)f_1(E_{\gamma},{\hat{T}}_c,q^2)+q^2{\cal{N}}_E^{(2)}f_2(E_{\gamma},{\hat{T}}_c,q^2)\Big]~,
\end{align}
where $\delta=\delta_0^1-\delta_0^2\simeq \pi/18$, $|{\cal{M}}_K|=1.81\times 10^{-8}$ GeV. As in~\cite{Cappiello:2011qc}, we have chosen to work with the set of variables $(E_{\gamma},{\hat{T}}_c, q)$, originally defined in~\cite{Christ:1967zz}, with the redefinition ${\hat{T}}_c=T_c+m_{\pi}$.  The kinematic functions $f_i$ are given by 
\begin{figure}[t]
\begin{center}
\includegraphics[width=8.0cm]{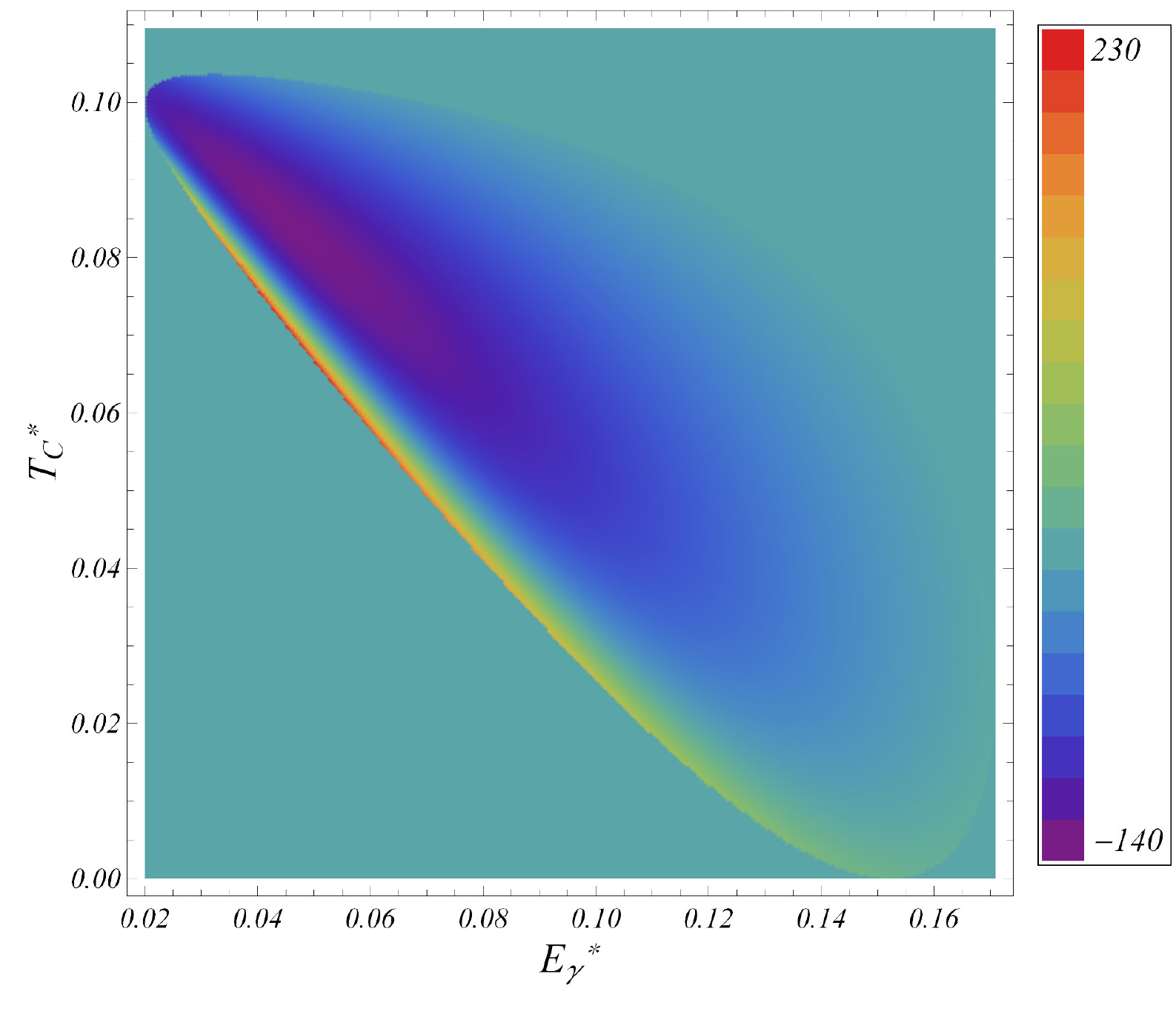}
\hskip 0.5cm
\includegraphics[width=8.0cm]{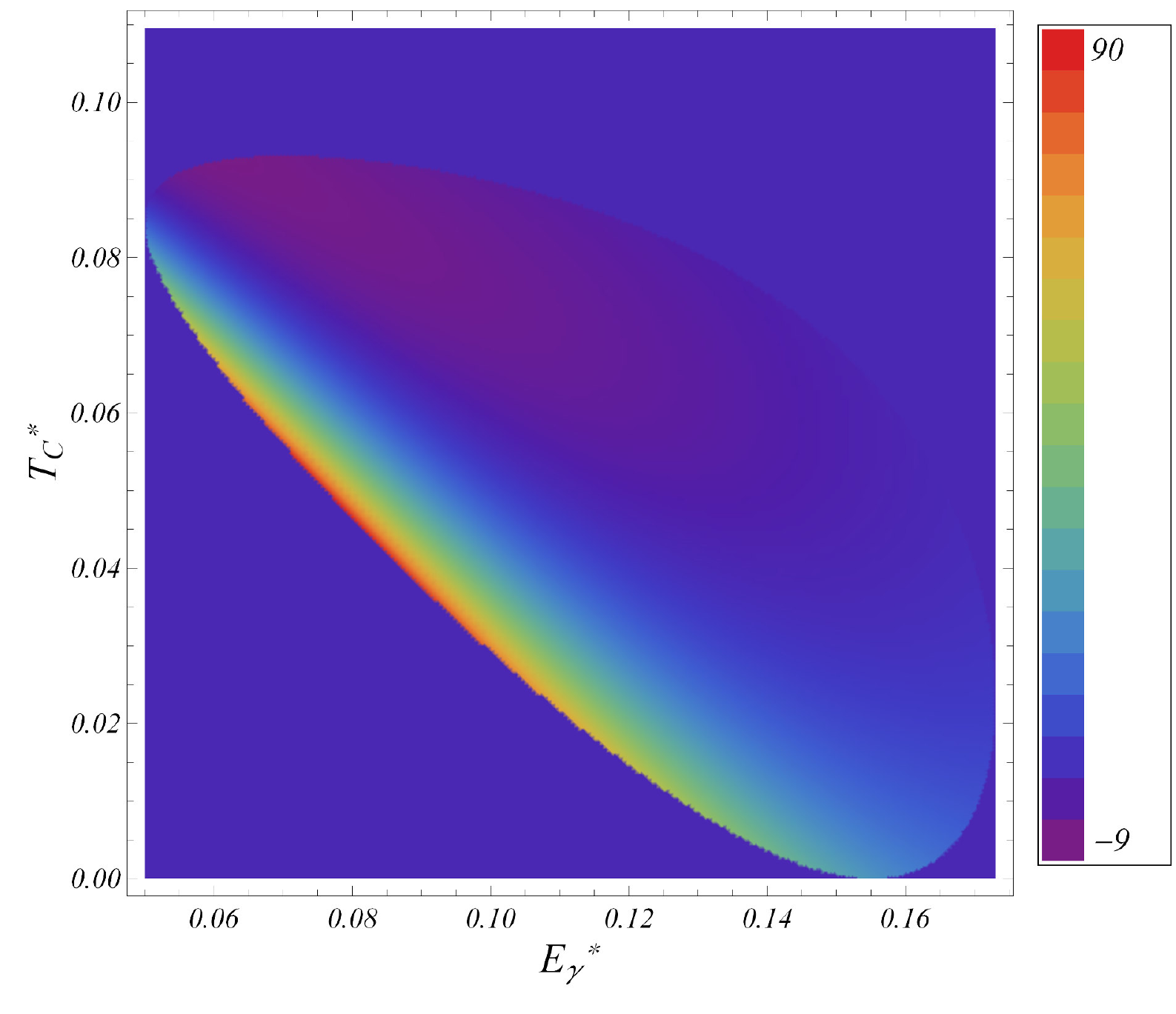}
\end{center}
\caption{\small{\it{Dalitz plots for the differential decay in the $(E_{\gamma},Tc)$ plane for $q=20$ MeV (left panel) and $q=50$ MeV (right panel). Numbers are given in units of $10^{-20}$ GeV$^{-1}$. The contour plot is 'spikier' the lower the $q$ values, a pattern mostly dictated by the structure of the Bremsstrahlung term.}}}\label{fig:1}
\end{figure}
\begin{align}
f_0(E_{\gamma}&,{\hat{T}}_c,q^2)=6m_K^6-24m_K^5(E_{\gamma}+{\hat{T}}_c)+3m_K^4\bigg[8E_{\gamma}^2+24E_{\gamma}{\hat{T}}_c+8{\hat{T}}_c^2+q^2\bigg]\nonumber\\
&-6m_K^3\bigg[8E_{\gamma}^2{\hat{T}}_c+E_{\gamma}(8{\hat{T}}_c^2-q^2)+3q^2{\hat{T}}_c\bigg]+6m_K^2\bigg[4E_{\gamma}^2(m_{\pi}^2-q^2)+2E_{\gamma}q^2{\hat{T}}_c+q^2(4{\hat{T}}_c^2-q^2))\bigg]\nonumber\\
&+6q^2m_K\bigg[E_{\gamma}(3q^2-6m_{\pi}^2)+q^2{\hat{T}}_c\bigg]+3q^4(4m_{\pi}^2-q^2);\nonumber\\
f_1(E_{\gamma}&,{\hat{T}}_c,q^2)=m_K^4-6E_{\gamma}m_K^3+2m_K^2\bigg[4E_{\gamma}^2+2E_{\gamma}{\hat{T}}_c-2{\hat{T}}_c^2+q^2\bigg]\nonumber\\
&+m_K\bigg[2E_{\gamma}(4m_{\pi}^2-3q^2)-4q^2{\hat{T}}_c\bigg]-q^2(4m_{\pi}^2-q^2);\nonumber\\
f_2(E_{\gamma}&,{\hat{T}}_c,q^2)=3m_K^4-4m_K^3\bigg[E_{\gamma}+2{\hat{T}}_c\bigg]+2m_K^2\bigg[4E_{\gamma}{\hat{T}}_c+2{\hat{T}}_c^2-q^2\bigg]\nonumber\\
&+4E_{\gamma}m_K\bigg[q^2-2m_{\pi}^2\bigg]+q^2(4m_{\pi}^2-q^2)~.
\end{align}
\begin{table}[t]
\begin{center}
\begin{tabular}{c|c|c|c|c|c|c}
\noalign{\vskip 2mm}
$q_c$ (MeV)& $10^8\times \Gamma_{\cal{B}}$ & $\displaystyle\left[\frac{\Gamma_{\cal{E}}}{\Gamma_{\cal{B}}}\right]^{-1}$&$\displaystyle\left[\frac{\Gamma_{\textsc{int}}}{\Gamma_{\cal{B}}}\right]^{-1}_{(1,1,1)}$&$\displaystyle\left[\frac{\Gamma_{\textsc{int}}}{\Gamma_{\cal{B}}}\right]^{-1}_{(1,0,1)}$&$\displaystyle\left[\frac{\Gamma_{\textsc{int}}}{\Gamma_{\cal{B}}}\right]^{-1}_{(1,1,0)}$&$\displaystyle\left[\frac{\Gamma_{\textsc{int}}}{\Gamma_{\cal{B}}}\right]^{-1}_{(0,1,1)}$\\
\noalign{\vskip 2mm}
\hline
\noalign{\vskip 2mm}
$2m_l$ & 418.27 & 1100& -253& -225& -115& 216\\
$2$ & 307.96 & 821& -265& -226& -98& 159\\
$4$ & 194.74& 529& -363& -264& -78& 101\\
$8$ & 109.60 & 304& 1587& -850& -59& 58\\
$15$ & 56.12 & 161& 102& 156& -43& 31\\
$35$ & 15.50 & 50& 18& 21& -26& 11\\
$55$ & 5.62 & 22& 7& 9& -18& 5\\
$85$ & 1.37 & 8& 3& 4& -13& 3\\
$100$ & 0.67 & 5& 2& 3& -11& 2\\
$120$ & 0.24 & 3& 1.6& 2& -10& 1.4\\
$140$ & 0.04 & 2& 1.0& 1.1& -8& 0.9\\
$180$ & 0.003 & 1& 0.7& 0.8& -7& 0.7 \\
\end{tabular}
\end{center}
\caption{\footnotesize{\it{Branching ratios for the Bremsstrahlung and the relative weight of the electric and electric interference terms for different cuts in $q$, starting at $q_{min}$ (first row) and ending at $180$ MeV. To highlight the role of the different counterterms, the last columns show how the interference term changes when they are switched off one at a time.}}}\label{tab:2}
\end{table}
Values for the counterterms ${\cal{N}}_E^{(0)}$ and ${\cal{N}}_E^{(1)}$ are taken from experiment (see the previous section).\footnote{Ref.~\cite{Cappiello:2011qc} used ${\cal{N}}_E^{(0)}=-0.0022$. The experimental value bears a positive sign, which we take in this note. All tables and plots in~\cite{Cappiello:2011qc} were consistently done with the negative choice for ${\cal{N}}_E^{(0)}$.} Using the experimental information of table~\ref{tab:1}, one gets the following prediction
\begin{align}\label{det}
{\cal{N}}_E^{(2)}=+0.089(11)+6N_{17}~,
\end{align}
which is entirely based on experimental input. The value and sign of $N_{17}$ is unknown and can only be determined from an actual measurement of ${\cal{N}}_E^{(2)}$. It is expected that $N_{17}$ be small (a number of theoretical models predict $N_{17}$ to be actually zero). In the following we will assume that $N_{17}$ is small enough that it can be neglected in eq.~(\ref{det}). Given the experimental error quoted in (\ref{det}), this is true if $|N_{17}|\lesssim 0.002$.  

In fig.~\ref{fig:1} we show the Dalitz plot for the differential decay rate in the $(E_{\gamma},T_c)$ plane for different values of the invariant dilepton mass $q$. As expected, the plots have a smoother behavior the larger $q$ is, since the Bremsstrahlung peak is located at small $q$. In table~\ref{tab:2} we provide the values of the interference term (compared to the Bremsstrahlung) for different choices of the counterterms at different cuts $q_c$. We use the shorthand notation 
\begin{align}
\left[\frac{\Gamma_{\textsc{int}}}{\Gamma_{\cal{B}}}\right]^{-1}_{({\cal{N}}_E^{(0)}, {\cal{N}}_E^{(1)}, {\cal{N}}_E^{(2)})}
\end{align}
where a subscript $(1,1,1)$ indicates that all counterterms take on their estimated values. When a counterterm is neglected, it is indicated by a nul value. Given the definition of the variable $q_c$, the first row corresponds to the fully integrated rate in $q^2$, while the subsequent rows cut the lower $q^2$ regions. 

Notice that ${\cal{N}}_E^{(0)}$ determines the overall sign of the interference term, while ${\cal{N}}_E^{(2)}$ is important to revert this sign at already low $q_c$ values. ${\cal{N}}_E^{(1)}$ has a marginal effect. It is interesting to note that a sizeable ${\cal{N}}_E^{(2)}$ tends to make the interference term smaller, though not dramatically. As the table shows, different patterns for the counterterm values lead to distinct behaviors of the interference term with $q_c$. This could be used to extract a value for ${\cal{N}}_E^{(2)}$ or, more generally, to fit for ${\cal{N}}_E^{(0)}$, ${\cal{N}}_E^{(1)}$ and ${\cal{N}}_E^{(2)}$.   
   
\begin{figure}[t]
\begin{center}
\includegraphics[width=7.8cm]{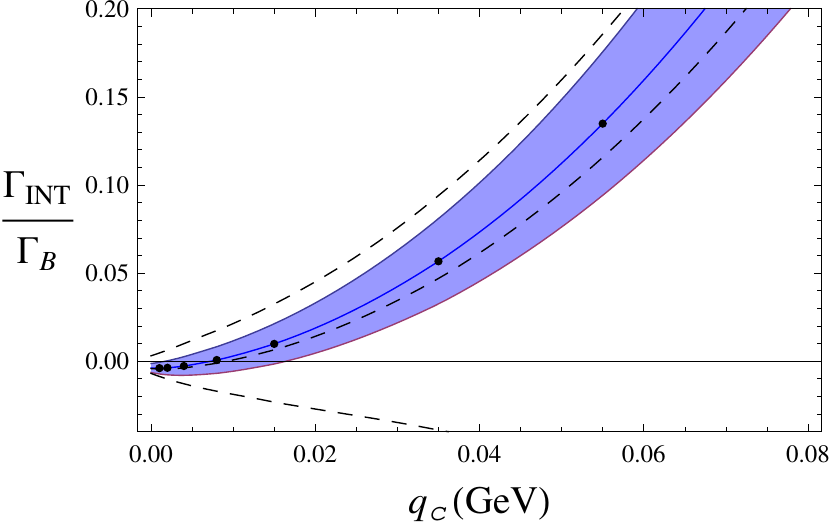}
\hskip 0.7cm
\includegraphics[width=7.8cm]{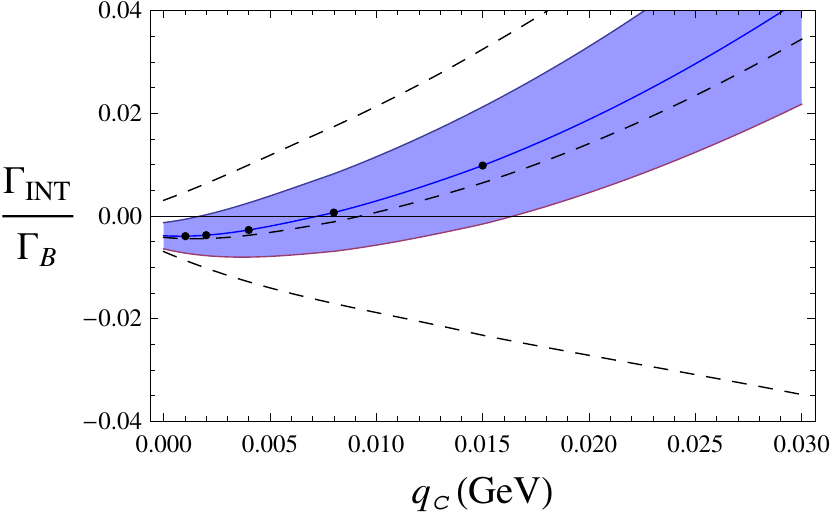}
\end{center}
\caption{\small{\it{Interference term normalized to the Bremstrahlung rate for different lower cuts in $q$. The points correspond to the (inverse) values shown in table~\ref{tab:2}. The band in blue corresponds to the 1 sigma level shift of the different counterterms. The dotted lines correspond to switching off the counterterms one at the time. The upper, central and lower lines corresponds to ${\cal{N}}_E^{(0)}=0$, ${\cal{N}}_E^{(1)}=0$ and ${\cal{N}}_E^{(2)}=0$, respectively. The right panel is a blown-up version of the left one for low $q_c$.}}}\label{fig:2}
\end{figure}

In fig.~\ref{fig:2} we plot the results of the last four columns of table~\ref{tab:2}, including also the error band, estimated by the 1 sigma shift of the experimental input. 
\section{Long-distance contributions to $K_S\to \pi^+\pi^-e^+e^-$}
The branching ratio for $K_S\to \pi^+\pi^-e^+e^-$ has already been measured by the NA48 collaboration~\cite{Lai:2000xf,Lai:2003ad}. In order to probe the chiral ${\cal{O}}(p^4)$ structure of this decay one needs to reach a percent precision, which might be possible at LHCb~\cite{MarinBenito:2017pcb}. Here we provide a first analysis of this decay mode, which bears parallelisms with $K_L\to \pi^+\pi^-e^+e^-$, already studied in~\cite{Elwood:1995xv,Pichl:2000ab}. 

A particularly interesting aspect of $K_S\to \pi^+\pi^-e^+e^-$ is that the magnetic piece is a CP-violating effect. This means that the interference term is, despite being suppressed, the relevant contribution after the Bremsstrahlung. The charge radius contribution can also be shown to be CP-violating. Thus, to a very good approximation, one finds that the hadronic form factors can be described by 
\begin{align}\label{result}
F_1&=\frac{2ie}{2q\cdot p_++q^2}{\cal{M}}_K-\frac{4ieG_8 e^{i\delta_1^1}}{f_{\pi}}\left[1-2\frac{m_K^2-m_{\pi}^2}{2q\cdot p_++q^2}\right]q^2L_9+\frac{2ieG_8 e^{i\delta_1^1}}{3f_{\pi}}\big\{q^2{\cal{N}}_E^{(3)}+6q\cdot p_-{\cal{N}}_E^{(0)}\big\}~,\nonumber\\
F_2&=-\frac{2ie}{2q\cdot p_-+q^2}{\cal{M}}_K+\frac{4ieG_8 e^{i\delta_1^1}}{f_{\pi}}\left[1-2\frac{m_K^2-m_{\pi}^2}{2q\cdot p_-+q^2}\right]q^2L_9-\frac{2ieG_8 e^{i\delta_1^1}}{3f_{\pi}}\big\{q^2{\cal{N}}_E^{(3)}+6q\cdot p_+{\cal{N}}_E^{(0)}\big\}~,\nonumber\\
F_3&=-\epsilon\frac{16eG_8 e^{i\delta_1^1}}{f_{\pi}}(N_{29}+N_{31})~,
\end{align}
where ${\cal{M}}_K$ is the matrix element of the $K_S\to\pi^+\pi^-$ decay, which can be written as
\begin{align}
{\cal{M}}_K=A_0e^{i\delta_0^0}+\frac{1}{\sqrt{2}} A_2e^{i\delta_0^2}\equiv |{\cal{M}}_K|e^{i\delta}~.
\end{align} 
Experimentally, one finds that $|{\cal{M}}_K|=3.92\times 10^{-7}$ GeV. 

The first term in both $F_1$ and $F_2$ is the Bremsstrahlung piece. In analogy with the analysis of the previous section, we evaluate it using Low's theorem, such that the important effect of final state interactions is taken into account. The remaining pieces in $F_1$ and $F_1$ collect the strong and weak ${\cal{O}}(p^4)$ corrections. In~(\ref{result}) we have used ${\cal{N}}_E^{(0)}=N_{14}-N_{15}-N_{16}-N_{17}$, which was already present in $K^+\to \pi^+\pi^0\gamma^*$, and defined
\begin{align}
{\cal{N}}_E^{(3)}&=N_{14}-N_{15}-3(N_{16}+N_{17})~,\nonumber\\
{\cal{N}}_E^{(4)}&=N_{14}-N_{15}-3(N_{16}-N_{17})~.
\end{align} 
The direct emission is estimated with the ${\cal{O}}(p^4)$ strong and weak counterterms, neglecting the loop corrections.

It is instructive to compare the results in eqs.~(\ref{result}) with the corresponding expressions for $K_L\to \pi^+\pi^-e^+e^-$ given in~\cite{Elwood:1995xv,Pichl:2000ab}: 
\begin{itemize}
\item In eq.~(\ref{result}) the Bremsstrahlung piece is the dominant contribution while the magnetic piece is a CP-violating effect. For $K_L$ one finds the reverse situation, which makes this latter decay mode especially suited to determine the magnetic counterterm $N_{19}+N_{31}$ combination. 

\item An interesting feature of eq.~(\ref{result}) is the $SU(3)$-violating pole structure that comes with $L_9$. In $K_L$ such a structure is also present but violates CP and is completely negligible. Such a term is also present in $K^+\to \pi^+\pi^0\gamma^*$, but there it represents a tiny ${\cal{O}}(m_{\pi^+}^2-m_{\pi^0}^2)$ isospin violation, which we neglected altogether in the previous section. In contrast, for $K_S\to \pi^+\pi^-\gamma^*$ it is a nonnegligible contribution.

\item The weak electric counterterm combinations are the same in both decay modes. However, ${\cal{N}}_E^{(3)}$ and ${\cal{N}}_E^{(4)}$ exchange their roles: for $K_S$ the impact of ${\cal{N}}_E^{(4)}$ is negligible, while for $K_L$ it is ${\cal{N}}_E^{(3)}$ that can be safely dismissed. 
\end{itemize}

The dominance of ${\cal{N}}_E^{(3)}$ over ${\cal{N}}_E^{(4)}$ is relevant. While the latter combination is presently unknown, the former can be actually predicted from the experimental values in table~\ref{tab:1} to be 
\begin{align}\label{pred}
{\cal{N}}_E^{(3)}=+0.040(5)~,
\end{align}
where the uncertainty is purely experimental. One therefore finds that the long-distance contributions to $K_S\to \pi^+\pi^-\gamma^*$ can be determined with remarkable accuracy, namely
\begin{align}
BR(K_S\to \pi^+\pi^-e^+e^-)=\underbrace{4.74\cdot 10^{-5}}_{\text{Brems.}}+\underbrace{4.39\cdot 10^{-8}}_{\text{Int.}}+\underbrace{1.33\cdot 10^{-10}}_{\text{DE}}~,
\end{align}
where in the interference term, by virtue of the $\Delta I=\tfrac{1}{2}$ rule, the relevant phase difference is to a very good approximation given by $\delta_0^0-\delta_1^1\simeq 34^0$~\cite{Colangelo:2001df}. This number is in excellent agreement with the PDG average~\cite{Patrignani:2016xqp}:
\begin{align}
BR(K_S\to \pi^+\pi^-e^+e^-)_{exp}=(4.79\pm 0.15)\times 10^{-5}~.
\end{align}
Similarly, one can predict that
\begin{align}
BR(K_S\to \pi^+\pi^-\mu^+\mu^-)=\underbrace{4.17\cdot 10^{-14}}_{\text{Brems.}}+\underbrace{4.98\cdot 10^{-15}}_{\text{Int.}}+\underbrace{2.17\cdot 10^{-16}}_{\text{DE}}~.
\end{align}

\begin{figure}[t]
\begin{center}
\includegraphics[width=8.0cm]{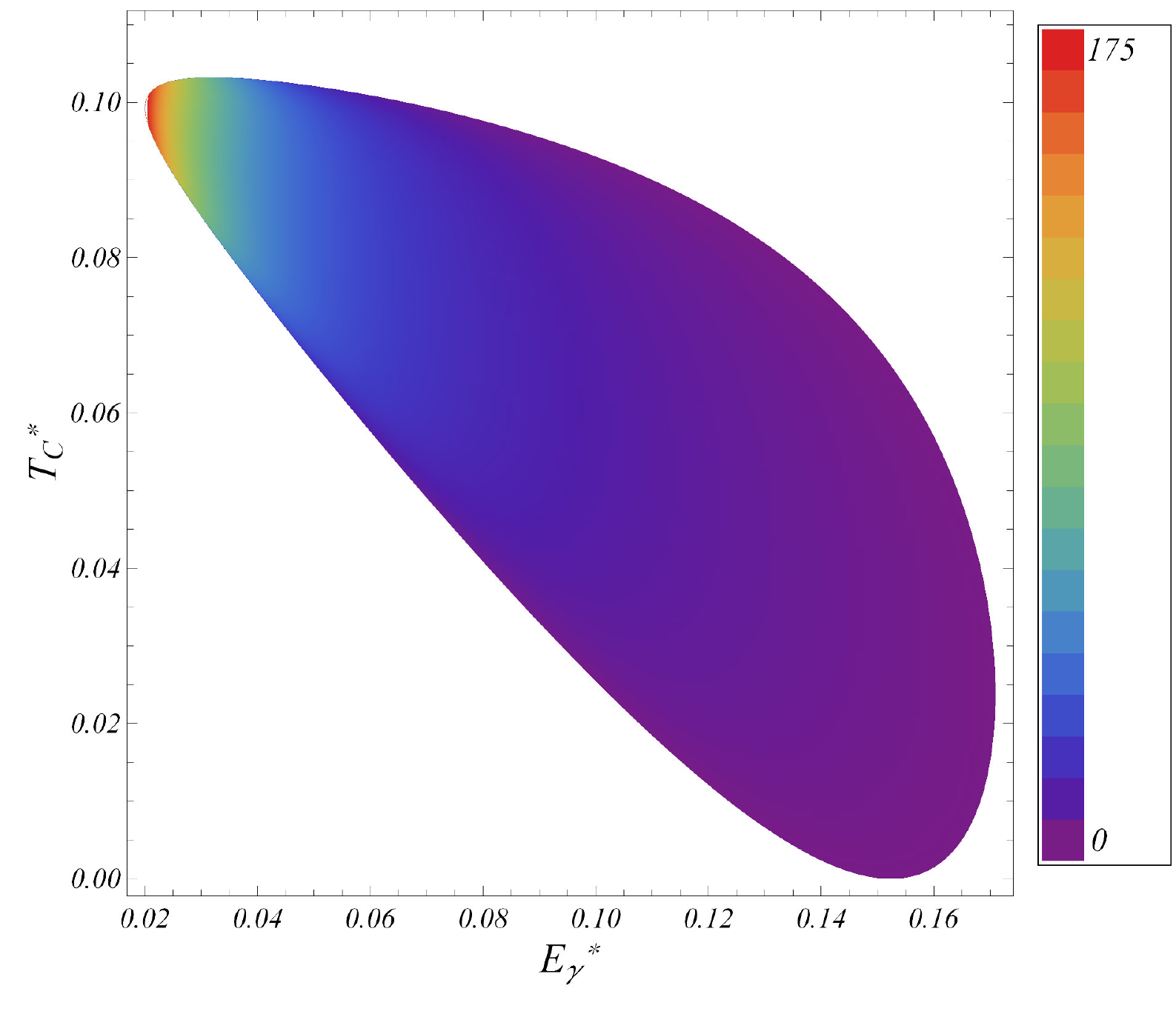}
\hskip 0.5cm
\includegraphics[width=8.0cm]{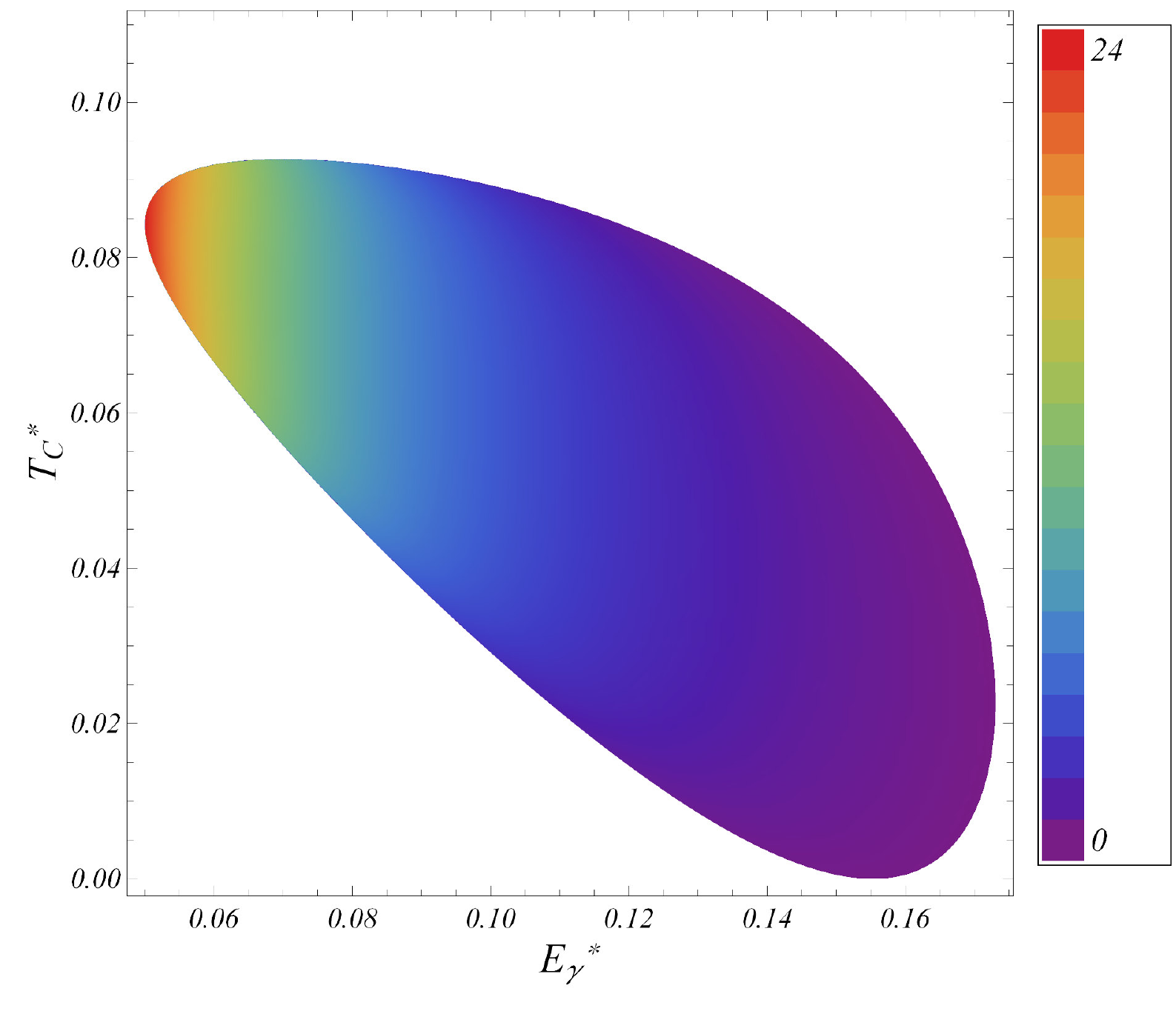}
\end{center}
\caption{\small{\it{Dalitz plots for the differential decay in the $(E_{\gamma},Tc)$ plane for $q=20$ MeV (left panel) and $q=50$ MeV (right panel). Numbers are given in units of $10^{-18}$ GeV$^{-1}$. The contour plot is 'spikier' the lower the $q$ values, a pattern mostly dictated by the structure of the Bremsstrahlung term.}}}\label{fig:5}
\end{figure}

Using the same formalism employed in the previous section for $K^+\to \pi^+\pi^0\gamma^*$, in fig.~\ref{fig:5} we show the Dalitz plot for $K_S\to \pi^+\pi^-e^+e^-$ in the $(E_{\gamma},T_c)$ plane for fixed dilepton invariant masses $q=20$ MeV and $q=50$ MeV. Note that, in contrast to $K^+\to\pi^+\pi^0e^+e^-$, the interference term is positive all over the physical region. 

\section{Concluding remarks}

\begin{figure}[t]
\begin{center}
\includegraphics[width=7.8cm]{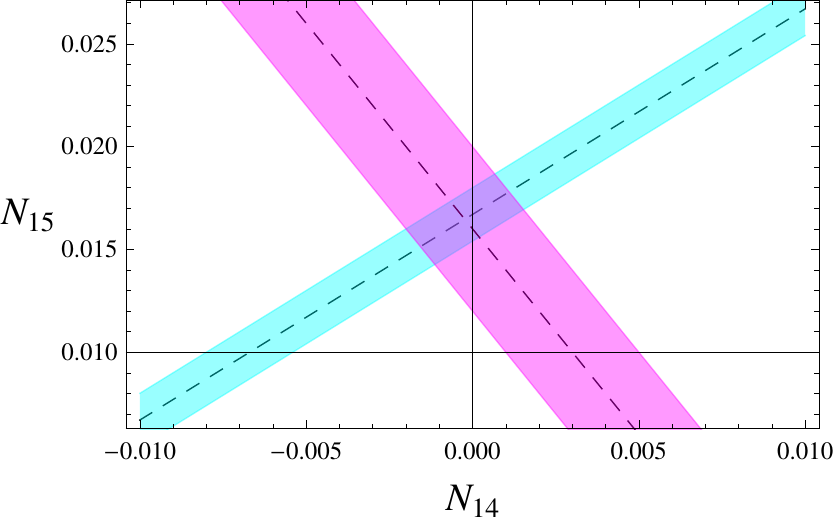}
\hskip 0.7cm
\includegraphics[width=7.8cm]{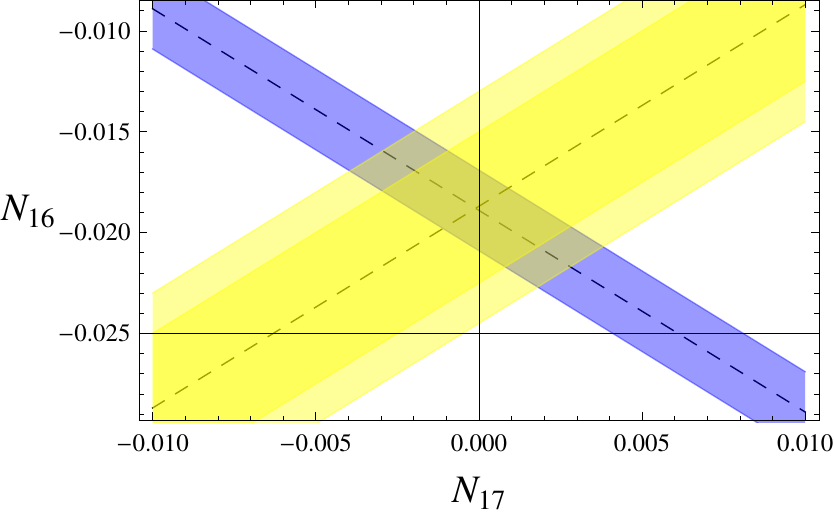}
\end{center}
\caption{\small{\it{Left panel: values of $N_{14}$ and $N_{15}$ as given by $K^{\pm}\to \pi^{\pm}\gamma^*$ (blue band) and $K_S\to \pi^{0}\gamma^*$ (violet band). Right panel: values for $N_{16}$ and $N_{17}$ extracted from $K^{\pm}\to \pi^{\pm}\pi^0\gamma$ (blue band) and $K^{\pm}\to \pi^{\pm}\pi^0e^+e^-$ (yellow band) measurements. The latter is an educated estimate (see main text).}}}\label{fig:4}
\end{figure}

The experimental situation in radiative kaon decays is at a point where the relevant (electric) weak chiral counterterms $N_{14}-N_{18}$ can be individually determined. The outcome of the measurements on $K^\pm\to \pi ^{\pm} \gamma^*$ and $K_{S}\to \pi ^{0} \gamma^*$ is plotted in fig.~\ref{fig:4} and gives 
\begin{align}
N_{14}&=(-2\pm 18)\times 10^{-4};\qquad N_{15}=(1.65\pm 0.22)\times 10^{-2}~.
\end{align}
Adding the measurements on $K^{\pm }\to \pi ^{\pm}\gamma\gamma$, one can also conclude that
\begin{align}
N_{18}=(-7.5\pm 2.3)\times 10^{-3}~.
\end{align}
The determination of $N_{16}$ and $N_{17}$ requires information on $K^{\pm }\to \pi ^{\pm }\pi ^{0}\gamma$ and $K^{\pm }\to \pi ^{\pm }\pi ^{0}\gamma^*$. The latter is currently being analyzed by the NA48/2 collaboration, which gathered 5000 events thereof. With this rather limited statistics, an extraction of $N_{16}$ and $N_{17}$ is only feasible if (a) there is some expectation for the size of the counterterm combinations to be tested; and (b) a strategy is devised to compensate the overwhelming dominance of the Bremsstrahlung contributions.

In this paper we have provided an estimate for the counterterms to be measured in $K^{\pm }\to \pi ^{\pm }\pi ^{0} e^+e^-$. Our estimates are, to the extent possible, based on the existing experimental knowledge of radiative kaon decays, and only very mild theoretical assumptions are used. We have shown that they generate a distinct shape in the Dalitz plot and, if judicious cuts on the dilepton invariant mass are used, one can largely compensate the relatively low statistics currently available. While the larger statistics of NA62 will definitely allow for a more precise determination of the counterterm combinations, the strategy proposed here is meant to make an extraction with NA48/2 data possible.

On the right panel of fig.~\ref{fig:4} we show estimates for $N_{16}$ and $N_{17}$, where the external yellow band corresponds to a value of $N_E^{(2)}=0.089(11)$, obtained assuming that $N_{17}$ is negligible. The internal yellow band corresponds to the same central value but assuming that a precision of $4\times 10^{-3}$ can be reached.  The experimental determination would fall out of the yellow band if $N_{17}$ turns out to be sizeable, roughly $|N_{17}|\gtrsim 2\times 10^{-3}$.

We will close this note with some comments on the theoretical understanding of the weak counterterm values, i.e., on the electroweak interactions at low energies. The situation is not as solid as for the strong sector, where the bulk of the nonperturbative effects entering the Gasser-Leutwyler coefficients $L_i$ is due to resonance exchange, with vector meson dominance (VDM) as a solid guiding principle. The weak counterterms, as opposed to the strong ones, are sensitive to the whole range of energies. Resonance models with VDM are therefore based on the assumption that the low-energy region is dominant. Given the large number of parameters to fit, additional simplifying assumptions are used in order to end up with predictive schemes. Failure to reproduce the experimental numbers can therefore be attributed to (a) a nonnegligible contribution from short distances inside the weak counterterms; and (b) assumptions on the long-distance contributions which are not supported phenomenologically.    

While the predictions of the different models should be taken with caution, they have merits that sometimes are not fully appreciated. In table~\ref{tab:3} we have listed a number of counterterm combinations relevant for different radiative kaon decays (see~\cite{DAmbrosio:1996jmq} for a more comprehensive list), together with the predictions of a number of models: the weak deformation model (WDM)~\cite{Pich:1990mw}, factorization model (FM)~\cite{Ecker:1990in}, holographic electroweak model (HEW)~\cite{Cappiello:2011re} and the resonance model studied in ref.~\cite{DAmbrosio:1997ctq}. 

\begin{table}[t]
\begin{center}
 \begin{tabular}{c|c|c|c } 
 \mbox{counterterm combinations}& \mbox{decay mode} & WDM/FM/HEW& $R^\mu$\\ \hline
$N_{14}-N_{15}$ &$K^{\pm}\to \pi^{\pm}\gamma^*$& $-3L_9-L_{10}-2H_1$ & $-0.020\eta_V+0.004\eta_A$\\ \hline
2$ N_{14}+N_{15} $&$K_S\to \pi^0\gamma^*$&$-2L_{10}-4H_1$ &$0.08\eta_V$\\ \hline
$N_{14}-N_{15}-N_{16}-N_{17}$&$K^{\pm}\to \pi^{\pm}\pi^0\gamma$ &$-2(L_9+L_{10})$ & $0.002\eta_V-0.010\eta_A$ \\ \hline 
$N_{14}-N_{15}-2N_{18}$&$K^{\pm}\to \pi^{\pm}\gamma\gamma$&$-3(L_9+L_{10})$&$-0.01\eta_A$ \\  \hline
$N_{14}+2N_{15}-3(N_{16}-N_{17})$&$K^{\pm}\to \pi^{\pm}\pi^0\gamma^*$&$6L_9-4L_{10}+4H_1$& $0.12\eta_V+0.01\eta_A$\\ \hline 
$N_{14}-N_{15}-3(N_{16}-N_{17})$&$K_L\to \pi^+\pi^-\gamma^*$&$-4L_{10}+4H_1$& $-0.004\eta_V+0.018\eta_A$\\ \hline 
$N_{14}-N_{15}-3(N_{16}+N_{17})$&$K_S\to \pi^+\pi^-\gamma^*$&$-4L_{10}+4H_1$& $0.05\eta_V-0.04\eta_A$\\ \hline 
$7(N_{14}-N_{16})+5(N_{15}+N_{17})$&$K_S\to \pi^+\pi^-\pi^0\gamma$&$10L_9-14L_{10}$& $0.4\eta_V+0.01\eta_A$
\end{tabular} 
\end{center}
\caption{\small{\it{Predictions of different counterterm combinations by two sets of models. On the third column we show the results of the weak deformation model (WDM), factorization model (FM) and holographic electroweak model (HEW), which yield the same predictions. On the fourth column, we list the estimates for the resonance model of ref.~\cite{DAmbrosio:1997ctq}, where numbers have been rounded up for internal consistency.}}}\label{tab:3}
\end{table}

The WDM/FM/HEW models express their predictions in terms of the strong counterterms $L_9$, $L_{10}$ and $H_1$. Leaving aside $H_1$, which has not been determined, the experimental values for $L_9$ and $L_{10}$ indicate that
\begin{align}\label{estimates}
N_{14}-N_{15}&\simeq +{\cal{O}}(10^{-2})~;\qquad & N_{14}+2N_{15}-3(N_{16}-N_{17})&\simeq +{\cal{O}}(10^{-2})~;\nonumber\\
2 N_{14}+N_{15}&\simeq -{\cal{O}}(10^{-2})~;\qquad & N_{14}-N_{15}-3(N_{16}-N_{17})&\simeq +{\cal{O}}(10^{-2})~;\nonumber\\
N_{14}-N_{15}-N_{16}-N_{17}&\simeq {\cal{O}}(10^{-3})~;\qquad & N_{14}-N_{15}-3(N_{16}+N_{17})&\simeq +{\cal{O}}(10^{-2})~; \nonumber\\
N_{14}-N_{15}-2N_{18}&\simeq {\cal{O}}(10^{-3})~;\qquad & 7(N_{14}-N_{16})+5(N_{15}+N_{17})&\simeq +{\cal{O}}(10^{-1})~.
\end{align} 
The counterterm combinations listed on the left have been actually measured (see table~\ref{tab:1}) and agree with the experimental numbers. Notice that the ${\cal{O}}(10^{-3})$ predictions are interpreted as accidental cancellations, proportional to $(L_9+L_{10})$, which is known to be a small number. Guessing the resulting sign when such cancellations occur is stretching the models beyond their range of applicability.\footnote{A similar situation is encountered in $K\to 3\pi$ decays. See ref.~\cite{Cappiello:2011re} for a detailed discussion.} The models also predict that the undetermined counterterms in $K^+\to\pi^+\pi^0e^+e^-$ and $K_S\to\pi^+\pi^-e^+e^-$ discussed in this paper should be large and positive.
     
In contrast to the WDM model, the resonance model of ref.~\cite{DAmbrosio:1997ctq} allows one to gauge the relative weight of vector and axial resonances as long-distance components of the weak chiral couplings and thereby test how robust is the VDM assumption. From table~\ref{tab:3} it is clear that if VMD is assumed, then there is a reasonable qualitative agreement with the conclusions of eq.~(\ref{estimates}). An exception is the counterterm entering $K_L\to \pi^+\pi^-\gamma^*$, which is predicted to have a marginal vectorial component and accordingly should be expected to be small. A determination of $N_{16}$ and $N_{17}$ through $K^+\to \pi^+\pi^0\gamma^*$ should resolve this issue.  
      
\section*{Acknowledgments}
We thank Brigitte Bloch-Devaux for very stimulating discussions and for her comments on a first version of this manuscript. O.C. thanks the University of Naples for a pleasant stay during the completion of this work. L.C. and G.D. were supported in part by MIUR under Project No. 2015P5SBHT and by the INFN research initiative ENP. The work of O.C. is supported in part by the Bundesministerium for Bildung und Forschung (BMBF FSP-105), and by the Deutsche Forschungsgemeinschaft (DFG FOR 1873). 


\end{document}